\documentclass[10pt,aps,prd,showpacs,superscriptaddress,twocolumn]{revtex4}
\usepackage{amsfonts}
\usepackage{amssymb}
\usepackage{amsmath}
\usepackage{graphicx}

\date{\today}

\begin{document}

\title{Instability of nonminimally coupled scalar fields 
in the spacetime of thin charged shells}

\author{J\'essica Santiago}\email{jessica.santiago@ufabc.edu.br}
\affiliation{Centro de Ci\^encias Naturais e Humanas,
Universidade Federal do ABC, 
Avenida dos Estados, 5001, 09210-580, 
Santo Andr\'e, S\~ao Paulo, Brazil}

\author{Andr\'e G.\ S.\ Landulfo}\email{andre.landulfo@ufabc.edu.br}
\affiliation{Centro de Ci\^encias Naturais e Humanas,
Universidade Federal do ABC, 
Avenida dos Estados, 5001, 09210-580,
Santo Andr\'e, S\~ao Paulo, Brazil}

\author{William C. C. Lima}
\email{william.correadelima@york.ac.uk}
\affiliation{Department of Mathematics, University of York, 
Heslington, York YO10 5DD, United Kingdom}

\author{George E. A. Matsas}
\email{matsas@ift.unesp.br}
\affiliation{Instituto de F\'\i sica Te\'orica, Universidade Estadual Paulista,
Rua Dr. Bento Teobaldo Ferraz 271, 01140-070, S\~ao Paulo, S\~ao Paulo, Brazil}

\author{Raissa F. P. Mendes}
\email{rmendes@uoguelph.ca}
\affiliation{Department of Physics, University of Guelph, Guelph, Ontario, 
N1G 2W1, Canada}

\author{Daniel A. T. Vanzella}
\email{vanzella@ifsc.usp.br}
\affiliation{Instituto de F\'\i sica de S\~ao Carlos,
Universidade de S\~ao Paulo, Caixa Postal 369, 13560-970, 
S\~ao Carlos, S\~ao Paulo, Brazil}

\begin{abstract}

We investigate the stability of a free scalar field nonminimally 
coupled to gravity under linear perturbations in the spacetime of 
a charged spherical shell. Our analysis is performed in the context 
of quantum field theory in curved spacetimes. This paper completes 
previous analyses which considered the exponential enhancement of 
vacuum fluctuations in the spacetime of massive shells. 

\end{abstract}


\maketitle

\section{Introduction}
\label{sec:introduction}

It has been shown that well-behaved spacetimes are able to induce an 
exponential enhancement of the vacuum fluctuations of some nonminimally 
coupled free scalar fields~\cite{lv}. This ``vacuum awakening effect'' 
may be seen as the quantum counterpart of the classical instability 
experienced by these fields under linear perturbations~\cite{mmv}. 
The exponential growth of the vacuum energy density of an unstable scalar 
field in the spacetime of, e.g., a neutron star~\cite{lmv} 
would necessarily induce the system to evolve into a new equilibrium 
configuration culminating in the emission of a burst of free scalar 
particles~\cite{llmv} (see also Refs.~\cite{novak98,pcbrs11,rdans12} 
for related classical analyses). Conversely, the determination 
of the mass-radius ratio of observed neutron stars may be used to rule out 
the existence of whole classes of nonminimally coupled scalar 
fields~\cite{m,pl}. This has motivated us to investigate how 
the vacuum awakening effect is impacted by relaxing some symmetries 
assumed in Ref.~\cite{lmv}. In order to avoid complications in modeling 
the fluid, the analyses of deviations from sphericity~\cite{lmmv} and 
staticity~\cite{mmv2} were considered in the context of massive thin 
shells. In this paper, we investigate the vacuum awakening mechanism 
when we endow a spherical shell with electrical charge. The presence
of charge affects both the energy conditions satisfied by the shell
matter and the effective potential appearing in the radial part of the 
scalar field equation. Hence, it is interesting to inquire how previous 
results for massive thin shells are modified in the charged case.

The paper is organized as follows. In Sec.~\ref{sec:spacetime}, 
we introduce the spacetime of a massive charged shell. 
In Sec.~\ref{sec:effect}, we quantize the real scalar field in 
this background and discuss the vacuum awakening effect.  
In Sec.~\ref{sec:vacuum}, we investigate the exponential growth of 
the vacuum energy density. Section~\ref{sec:finalremarks} is dedicated
to our final remarks. We assume metric signature $(- + + +)$ and natural 
units in which $c=\hbar=G=1$ unless stated otherwise.

\section{Spherically symmetric charged shells}
\label{sec:spacetime}

Let us write the line element describing the spacetime of a spherically symmetric 
thin shell with mass $M$ and electric charge $Q$ lying at the radial coordinate 
$r = \mathsf{R}$
as~\cite{b}
\begin{equation}
	ds^2_- = - f(\mathsf{R}) dt^2 + dr^2  + r^2 \left( d\theta^2 + \sin^2\theta d\varphi^2 \right)
\label{ds_in}
\end{equation}
and
\begin{equation}
		ds^2_+ = - f(r) dt^2 + f(r)^{-1} dr^2  + r^2 
		\left( d\theta^2 + \sin^2\theta d\varphi^2 \right)
\label{ds_out}
\end{equation}
with
\begin{equation}
f(r) \equiv (1-2M/r+Q^2/r^2) > 0,
\label{f(r)}
\end{equation}
where $\mathsf{R} > M + \sqrt{M^2-Q^2}$ when $|Q|\leq M$ (while $\mathsf{R}$ 
may assume any positive value when $|Q|>M$) and $\mp$ labels quantities defined 
at $r \lessgtr \mathsf{R}$, respectively. The three-dimensional timelike 
surface $\cal{S}$ defined 
at $r = \mathsf{R}$ will be covered with coordinates 
$\zeta^a = (t, \theta, \varphi)$. The metrics 
$h_{ab}^-$ and $h_{ab}^+$ on $\cal{S}$ as induced from the internal- 
and external-to-the-shell spacetime portions, respectively, satisfy 
$h_{ab}^- = h_{ab}^+$ as demanded by the continuity condition. 

The shell stress-energy-momentum tensor   
\begin{equation}
	T^{\mu\nu} = S^{ab} e^\mu_a e^\nu_b \delta(\ell)
\label{Tmunu}	
\end{equation}
can be computed from the discontinuity of the extrinsic 
curvature $\Delta K^{ab}\equiv K^{ab}_+ - K^{ab}_-$ across 
$\cal{S}$~(see, e.g., Ref.~\cite{poisson}). 
In Eq.~(\ref{Tmunu}), 
\begin{equation}
	S^{ab} \equiv -\frac{1}{8\pi}(\Delta K^{ab} - h^{ab} \Delta K),
\end{equation}
where $\Delta K \equiv \Delta K_{ab} h^{ab}$,
$e_a^\mu \equiv \partial x^\mu/\partial \zeta^a$ are the components 
of the coordinate vectors $\partial/\partial \zeta^a$ defined on 
$\cal{S}$, 
and $\ell$ (inside the delta distribution) is the proper distance along geodesics 
intercepting orthogonally $\cal{S}$ (with $\ell<0$, $\ell=0$, and $\ell>0$  
inside, on, and outside $\cal{S}$, respectively). By using 
Eqs.~(\ref{ds_in}) and (\ref{ds_out}), we obtain
\begin{equation}
S_{\;\;0}^0 = \frac{1}{4\pi \mathsf{R}} \left( f(\mathsf{R})^{1/2} -1 \right)
\label{S00}
\end{equation}
and
\begin{equation}
S_{\;\;2}^2  = S_{\;\;3}^3 
             = \frac{1}{8\pi \mathsf{R}} 
	               \left[ 
	               \left( 1-\frac{M}{\mathsf{R}} \right) f(\mathsf{R})^{-1/2} - 1 
	               \right]. 
\label{S22} 
\end{equation}
 
\begin{figure}
\includegraphics[scale=.25]{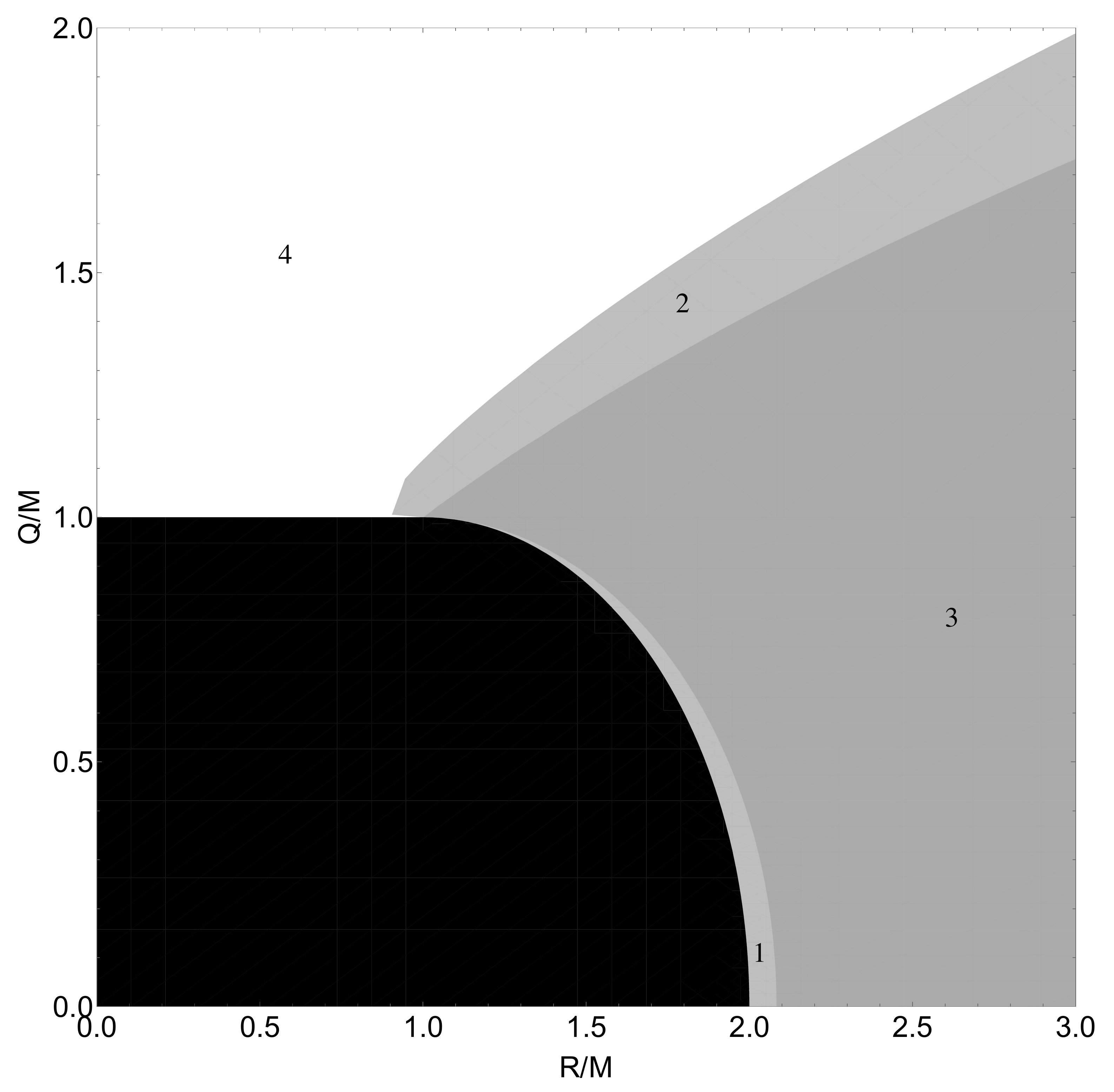}
\caption{The graph shows where the different energy 
conditions are satisfied as functions of the shell parameters 
$\mathsf{R}/M$ and $Q/M$ out of the black area. The black area
is excluded because there are no static shell configurations
in this region. The weak energy condition is satisfied everywhere 
but in region~4. The 
strong energy condition is violated in regions~2 and~4, 
while the dominant energy condition is violated in regions~1
and~4. We see, thus, that in a large portion, namely, region~3,
the three energy conditions are satisfied.}
\label{fig1:EC}
\end{figure}

In order to unveil the most realistic shell 
configurations, we have investigated the behavior of the 
different energy conditions with respect to the shell 
parameters $\mathsf{R}/M$ and $Q/M$. (For the stability 
of charged shells under linear perturbations, see 
Ref.~\cite{es}.)  We can see that the weak, 
strong, and dominant energy conditions are simultaneously 
satisfied in a large portion of Fig.~\ref{fig1:EC}, namely, 
region~3. 

\section{Quantizing the field and awaking the vacuum}
\label{sec:effect}

Now, we consider a nonminimally coupled massless real scalar 
field $\Phi$ satisfying the Klein-Gordon equation
\begin{equation}
	-\nabla_\mu\nabla^\mu\Phi+\xi R\Phi=0
\label{KG_eq}
\end{equation}
in the spacetime of our spherically symmetric charged 
shell, where $\xi  \in \mathbb{R}$ 
and $R$ is the scalar curvature. 
We follow the canonical procedure and expand the 
corresponding field operator~\cite{fulling,wald_qftcs}
\begin{equation}
\hat \Phi 
= 
\int d\vartheta (\eta) 
[\hat a_\eta u_\eta +\hat a_\eta^\dagger u_\eta^{*}],
\label{Phi_expandido}
\end{equation}
in terms of positive, $u_\eta$, and negative, 
$u_\eta^{*}$, norm modes with respect to the 
Klein-Gordon inner product, where $\vartheta$ 
is a measure defined on the set of quantum 
numbers $\eta$. The annihilation 
$\hat a_\eta$ and creation $\hat a_\eta^\dagger$
operators satisfy 
$[\hat a_\alpha, \hat a_\beta^\dagger] = \delta (\alpha, \beta)$
and
$[\hat a_\alpha, \hat a_\beta] = 0$, where the delta function
is defined by 
$\int d\vartheta (\alpha) \delta ( \alpha,\beta) \mathcal{F} (\alpha) 
= \mathcal{F} (\beta)$ 
for $\mathcal{F} \in C^\infty$, and the vacuum state 
$| 0 \rangle$ must satisfy $\hat a_\eta |0\rangle = 0$ 
for all $\eta$.    

The spacetime symmetries drive us to look for positive-norm 
modes in the form
\begin{equation}\label{mode_form}
	u^\mp_{\sigma l\mu} (t, r, \theta, \varphi) 
	= 
	T_\sigma(t) 
	F^\mp_{\sigma l}(r)	 
	Y_{l\mu} (\theta, \varphi),
\end{equation}
where $Y_{l\mu} (\theta, \varphi)$ represent the spherical 
harmonic  functions, $l=0,1,2 \ldots$, $\mu= -l, -l+1,\ldots l$, 
$\sigma = {\rm const} \in \mathbb{R}$, and 
\begin{equation}
i\left (T^*_\sigma \frac{dT_\sigma}{dt} - T_\sigma \frac{dT^*_\sigma}{dt} \right)>0
\label{positive-norm condition}
\end{equation}
(positive-norm condition). By using Eq.~(\ref{mode_form}) in 
Eq.~(\ref{KG_eq}), we obtain that  $F^\mp_{\sigma l} (r)$ and $T_\sigma (t)$
satisfy
\begin{equation}\label{F-_eq}
	-\frac{f(\mathsf{R})}{r^2} \frac{d}{dr} \left( r^2 \frac{d F^-_{\sigma l} }{dr} \right)
	+\frac{f(\mathsf{R})}{r^2} l(l+1) F^-_{\sigma l} = \sigma F^-_{\sigma l},
\end{equation}
\begin{equation}\label{F+_eq}
	-\frac{f(r)}{r^2} \frac{d}{dr} \left( r^2 f(r) \frac{d F^+_{\sigma l} }{dr} \right)
	+\frac{f(r)}{r^2} l(l+1) F^+_{\sigma l} = \sigma F^+_{\sigma l},
\end{equation}
and
\begin{equation}\label{T_eq}
	\frac{d^2}{dt^2}T_\sigma + \sigma T_\sigma = 0.
\end{equation}
Depending on the sign of $\sigma$, Eq.~(\ref{T_eq}) supplemented 
by condition~(\ref{positive-norm condition}) admits two kinds of general 
solutions:
\begin{equation}\label{T_osc_form}
 T_\sigma(t) \stackrel{\sigma = \omega^2}{\propto} \exp (-i \omega t),\;\;\; \omega >0,
\end{equation}
for $\sigma \equiv \omega^2>0$ and
\begin{equation}\label{T_exp_form}
 T_\sigma(t) \stackrel{\sigma = -\Omega^2}{\propto} e^{\Omega t-i\pi/12}+e^{-\Omega
t+i\pi/12},\;\;\; \Omega >0,
\end{equation}
for $\sigma\equiv -\Omega^2<0$. The combination chosen on the 
right-hand side of Eq.~(\ref{T_exp_form}) guarantees that the
corresponding $u^\mp_{\sigma l\mu} $ functions will be
indeed positive-norm modes~\cite{lv}. The modes~(\ref{mode_form})
with $T_\sigma (t)$ as given by Eqs.~(\ref{T_osc_form}) 
and~(\ref{T_exp_form}) will be denoted as 
\begin{equation}\label{mode_form-spherical1}
u^\mp_{\sigma l\mu} \stackrel{\sigma= \omega^2}{\longrightarrow}	
v^\mp_{\omega l \mu} = \frac{\psi^\mp_{\omega l}(r)}{r}
	Y_{l\mu} (\theta, \varphi) \frac{\exp (-i \omega t)}{\sqrt{2\omega}}
\end{equation}
and 
\begin{equation}\label{mode_form-spherical2}
u^\mp_{\sigma l\mu} \stackrel{\sigma= -\Omega^2}{\longrightarrow} 
w^\mp_{\Omega l \mu} =  \frac{\psi^\mp_{\Omega l}(r)}{r}
Y_{l\mu} (\theta, \varphi)  \frac{(e^{\Omega t-i\pi/12}+e^{-\Omega
t+i\pi/12})}{\sqrt{2\Omega}},
\end{equation}
corresponding to the usual time-oscillating 
and ``tachyonic" modes, respectively. 

The field-operator expansion~(\ref{Phi_expandido}) can be cast,
then, in terms of oscillatory and tachyonic modes 
as
\begin{align}
\hat \Phi 
& = 
\sum_{l\mu} \int d\omega    
[\hat b_{\omega l \mu} v_{\omega l \mu} + \hat b_{\omega l \mu}^\dagger v^*_{\omega l \mu}]
\nonumber \\
&+ 
\sum_{l\mu \Omega}    
[\hat c_{\Omega l \mu} w_{\Omega l \mu} + \hat c_{\Omega l \mu}^\dagger w^*_{\Omega l \mu}],
\label{Phi_expandido_spherical}
\end{align}
where the only nonzero commutation relations between the creation and annihilation operators 
are
\begin{align}
[\hat b_{\omega l \mu}, \hat b^\dagger_{\omega' l' \mu'}] 
& =  \delta_{ll'} \delta_{\mu \mu'}\delta (\omega-\omega'),
\\
[\hat c_{\Omega l \mu}, \hat c^\dagger_{\Omega' l' \mu'}] 
& = \delta_{ll'} \delta_{\mu \mu'}\delta_{\Omega \Omega'}.
\end{align}
The presence of tachyonic modes makes the scalar field unstable. 
Quantum fluctuations and, consequently, the expectation value of 
the stress-energy-momentum tensor, grow exponentially in 
time in the presence of these modes (see also Refs.~\cite{l13,ss70}
for further discussions on the quantization of unstable 
linear fields in static globally hyperbolic spacetimes).

Now, we shall discuss in detail the radial part of modes
$v^\mp_{\omega l\mu}$ and $w^\mp_{\Omega l\mu}$, namely, 
$\psi^\mp_{\omega l}(r)/r$ and
$\psi^\mp_{\Omega l}(r)/r$, respectively. For the sake of 
convenience, we define the coordinates 
$\chi_{\mp} =\chi_{\mp}(r)$  such that for $r<\mathsf{R}$ 
\begin{equation}
r \to 
\chi_{-}(r)  \equiv 
\frac{r}{f(\mathsf{R})^{1/2}}, 
\;\; {\rm for}\;\;  {\rm all} \;\; Q  \in \mathbb{R},
\label {chi-}
\end{equation}
while for $r \geq \mathsf{R}$
\begin{eqnarray}
\!\!\!\!\!r \to 
\chi_{+} (r) 
& \equiv &  
r  + \frac{\mathsf{R}_{2}^2 
\ln (r-\mathsf{R}_{2})-\mathsf{R}_{1}^2 
\ln (r-\mathsf{R}_{1})}{\mathsf{R}_{2}-\mathsf{R}_{1}}
\nonumber \\
& + & D^{(1)}, 
\;\; {\rm for}\;\; |Q|  < M,
\label {chi+2}
\end{eqnarray}
\begin{eqnarray}
\!\!\!\!\! r \to
\chi_{+}   (r) 
 &\equiv & 
r  - \frac{M^2}{r-M} + 2M \ln (r-M) 
\nonumber \\
& + & D^{(2)}, 
\;\; {\rm for}\;\; |Q| = M,
\label {chi+3}
\end{eqnarray}
and
\begin{eqnarray}
\!\!\!\!\! r \to 
\chi_{+}  (r)
&\equiv  &  r + \frac{2M^2-Q^2}{\sqrt{Q^2-M^2}} 
  \arctan \left[ \frac{r-M}{\sqrt{Q^2-M^2}} \right]  
\nonumber \\
& + & M\ln[r^2 f(r)]
+ D^{(3)}, 
\;\; {\rm for}\;\; |Q|  > M,
\label {chi+1}
\end{eqnarray}
where
$\mathsf{R}_{1}\equiv M-\sqrt{M^2-Q^2}$,
$\mathsf{R}_{2}\equiv M+\sqrt{M^2-Q^2}$,
and
$D^{(1)},D^{(2)},$ and $D^{(3)}$ are constants chosen such that 
$\chi_{-}$ and $\chi_{+}$ fit each other continuously on the shell. 
In terms of the coordinates $\chi_{\mp}= \chi_{\mp}(r)$, 
Eqs.~(\ref{F-_eq}) and (\ref{F+_eq}) can be rewritten as 
``Schr\"odinger-like" equations for $\psi_{\lambda l}^{\mp}$
where $\lambda \in \{\omega, \Omega\}$:
\begin{equation}\label{schrodinger_type_eq1}
- d^2 \psi_{\omega l}^{\mp}/ d\chi^2_{\mp} +V_{\textrm{eff}}^{(l,\mp)}\psi_{\omega l}^{\mp}
=\omega^2 \psi_{\omega l}^{\mp}
\end{equation}
and
\begin{equation}\label{schrodinger_type_eq2}
- d^2 \psi_{\Omega l}^{\mp}/ d\chi^2_{\mp} +V_{\textrm{eff}}^{(l,\mp)}\psi_{\Omega l}^{\mp}
=-\Omega^2 \psi_{\Omega l}^{\mp}
\end{equation}
with
\begin{equation}
V_{\textrm{eff}}^{(l,-)} (r) = f(\mathsf{R}) \frac{l(l+1)}{r^2}
\label{V-}
\end{equation}
and
\begin{equation}
V_{\textrm{eff}}^{(l,+)} (r) = f(r)
                           \left(\frac{2M}{r^3}- \frac{2 Q^2}{r^4}+ \frac{l(l+1)}{r^2}\right).
\label{V+}
\end{equation}
We note that the discontinuity of the potential across the shell is 
$
\Delta V_{\textrm{eff}} =  
f(\mathsf{R}) ( 2M/\mathsf{R}^3 -2Q^2/\mathsf{R}^4)
$
and that solutions $\psi_{\Omega l}^{+}(r)$ must vanish asymptotically 
to be normalizable:
\begin{equation}
\lim_{r \to +\infty} \psi^+_{\Omega l } = 0.
\label{psi_at_infinity}
\end{equation}

Now, let us discuss the junction conditions on $\psi_{\lambda l}^{\mp}$
for $\lambda \in \{\omega, \Omega\}$. 
By using Einstein equations in conjunction with Eq.~(\ref{Tmunu}), 
we have that
\begin{equation}
R = - 8\pi T = -8 \pi S \delta (\ell),
\label{R}
\end{equation}
where, by using Eqs.~(\ref{S00}) and~(\ref{S22}), we have
$$
S\equiv S^{ab} h_{ab}=\frac{1}{4\pi \mathsf{R} } 
\left[
\frac{2- 3 M/\mathsf{R} + Q^2/\mathsf{R}^2}{f(\mathsf{R})^{1/2}} - 2
\right].
$$
By using Eq.~(\ref{R}) in Eq.~(\ref{KG_eq}), we obtain the
junction conditions for $\psi^\mp_{\lambda l}(r)/r$ and 
for its first derivative 
along the direction orthogonal to $\cal{S}$, namely, 
\begin{equation} \label{continuity}
	[\psi^+_{\lambda l}/r - \psi^-_{\lambda l}/r]_{\cal{S}}
	 	= 0,
\end{equation}
and 
\begin{equation} \label{discontinuity} 
	\left[ \frac{d(\psi^+_{\lambda l}/r)}{d \ell} 
       - \frac{d(\psi^-_{\lambda l}/r)}{d \ell}
\right]_{\cal{S}} 
	=  \xi \gamma  (\psi_{\lambda l}/r)_{\cal{S}},
\end{equation}
where $\gamma = -2 \Delta K = -8\pi S$ and 
$(\psi_{\lambda l}/r)_{\cal{S}}= (\psi^\mp_{\lambda l}/r)_{\cal{S}}$
because of Eq.~(\ref{continuity}). 

Next, we shall look for the conditions on $M/\mathsf{R}$, $Q/M$, 
and $\xi$ which allow for the existence of solutions 
$\psi^\mp_{\Omega l}$ for Eq.~(\ref{schrodinger_type_eq2})
associated with regular modes $w_{\Omega l \mu}^\mp$ [see
Eq.~(\ref{mode_form-spherical2})]. The regularity requirement 
demands that $ \psi^-_{\Omega l }$ vanishes at the origin. The fact that 
the Klein-Gordon inner product fixes the mode normalization up to a 
multiplicative phase, allows us to assume without loss of generality 
that $ \psi^-_{\Omega l }$ approaches zero from positive values:
\begin{equation}
\lim_{r \to 0} \psi^-_{\Omega l } = 0^+.
\label{psi_at_origin}
\end{equation}
Then, by using Eqs.~(\ref{schrodinger_type_eq2})  
and~(\ref{psi_at_origin}) together with the 
junction condition~(\ref{continuity}) and the fact
that $V_{\rm eff}^{(l,-)} \geq 0$ [see  Eq.~(\ref{V-})],
we conclude that   
\begin{equation}
0<\psi^-_{\Omega l }|_{\cal S} = \psi^+_{\Omega l }|_{\cal S},
\label{continuity_spherical}
\end{equation}
while we see from Eq.~(\ref{discontinuity}) that in general
$d\psi^-_{\Omega l }/dr|_{\cal S} \neq d\psi^+_{\Omega l }/dr|_{\cal S}$.

From the usual analysis of bound states in nonrelativistic quantum mechanics, 
it is straightforward to infer that for a fixed $l$ the $\psi^\mp_{\Omega l }$ 
solution of Eq.~(\ref{schrodinger_type_eq2}) (if any) satisfying 
conditions~(\ref{psi_at_infinity}) and~(\ref{continuity})--(\ref{psi_at_origin}) 
with the $n${\em th} largest $\Omega$  will possess $n-1$ zeros for $r>0$. 
In particular, for those configurations admitting one single $\psi^\mp_{\Omega l }$ 
solution, we will have $\psi^\mp _{\Omega l } \geq 0$ (with the equality 
holding at the origin). The boundary separating configurations possessing 
and not possessing tachyonic modes for a fixed $l$ will be given 
by the set $M/\mathsf{R}$, $Q/M$, and $\xi$, which allows for a single 
well-behaved solution $\psi^\mp_{\Omega l }$  with $\Omega=0$.
These {\em marginal solutions} $\psi^\mp_{0 l }$ will also satisfy conditions~(\ref{continuity})--(\ref{psi_at_origin}). 
The asymptotic behavior of marginal solutions can be obtained 
directly from Eq.~(\ref{F+_eq}):
\begin{equation}
- \frac{d}{dr} \left( r^2 \frac{d \psi^+_{0 l}/r }{dr} \right)
	+ l(l+1) \frac{\psi^+_{0 l}}{r} \stackrel{r \gg M,Q}{\approx}  0,
\label{aux2}
\end{equation}
whose general solution can be cast as 
\begin{equation} 
	\psi^+_{0 l} ({r}) \stackrel{r \gg M,Q}{\approx}  B_l {r}^{l+1} +C_lr^{-l}
	\label{aux3}
\end{equation}
with $B_l, C_l \in \mathbb{R}$ being constants. We see that, for $l=0$, 
condition~(\ref{psi_at_infinity}) is actually not satisfied  
by any nontrivial $\psi_{0 0}^+$.
This is an artifact which appears because marginal solutions have $\Omega=0$. 
The presence of any small $\Omega \neq 0$ would eventually dominate for large 
enough $r$ changing the asymptotic form of Eq.~(\ref{aux2}) to
\begin{equation}
\frac{d}{dr} \left( r^2 \frac{d \psi^+_{\Omega l}/r }{dr} \right)
	- l(l+1) \frac{\psi^+_{0 l}}{r}
\stackrel{r \gg M,Q}{\approx} 
r \Omega^2 \psi_{\Omega l}^+,
\end{equation}
in which case the general solution could be written as 
\begin{equation} 
	\psi^+_{\Omega l} ({r}) \stackrel{r \gg M,Q}{\approx}  
	B_l {\cal{J}}_{\Omega l}^{(1)} (r)
+ C_l {\cal{J}}_{\Omega l}^{(2)} (r),
	\label{aux4}
\end{equation}
where for  small enough $\Omega$
$$
{\cal{J}}_{\Omega l}^{(1)} 
\stackrel{ M,Q \ll r \ll 1/\Omega}{\sim} r^{l+1},\;\;\; 
{\cal{J}}_{\Omega l}^{(1)} 
\stackrel{r \gg 1/\Omega}{\sim} \exp(\Omega r),
$$
$$
{\cal{J}}_{\Omega l}^{(2)} 
\stackrel{ M,Q \ll r \ll 1/\Omega}{\sim} r^{-l},\;\;\;
{\cal{J}}_{\Omega l}^{(2)} 
\stackrel{r \gg 1/\Omega}{\sim} \exp(-\Omega r).
$$
The fact that Eq.~(\ref{psi_at_infinity}) demands $B_l=0$ 
implies that we should look for marginal solutions~(\ref{aux3}) 
behaving asymptotically as $\psi^+_{0 l} ({r}) \stackrel{r \gg M,Q}{\sim}  r^{-l}$.    

It is easy to see that 
\begin{equation} \label{solution_in}
	\psi^-_{0 l} ({r})  = A_l {r}^{l+1}/\mathsf{R}^l, \;\;A_l>0,
\end{equation}
is a marginal solution in the {\em interior} of the shell 
satisfying condition~(\ref{psi_at_origin}) 
for any charge value. Now, we note from  
Eqs.~(\ref{V-})-(\ref{V+}) that the smaller the $l$ the ``deeper" 
the effective potential $V^{(l,\mp)}_{\rm eff}$ making tachyonic 
modes with vanishing angular momentum the most likely ones to exist. 
Because the appearance of a single tachyonic mode is enough to render
the system unstable, we shall restrict our quest to marginal solutions with
$l=0$. Then, from Eq.~(\ref{solution_in}), we immediately write 
\begin{equation} \label{solution_in,l=0}
	\psi^-_{0 0} ({r})  = A_0 r, \;\;A_0>0
\end{equation}
for all $Q$. Next, we look for marginal solutions 
{\em external} to the shell. For the sake of clarity, we present 
the cases $|Q| < M$, $|Q| = M$, and $|Q| > M$, separately. 

\begin{figure}[t]
\includegraphics[scale=0.25]{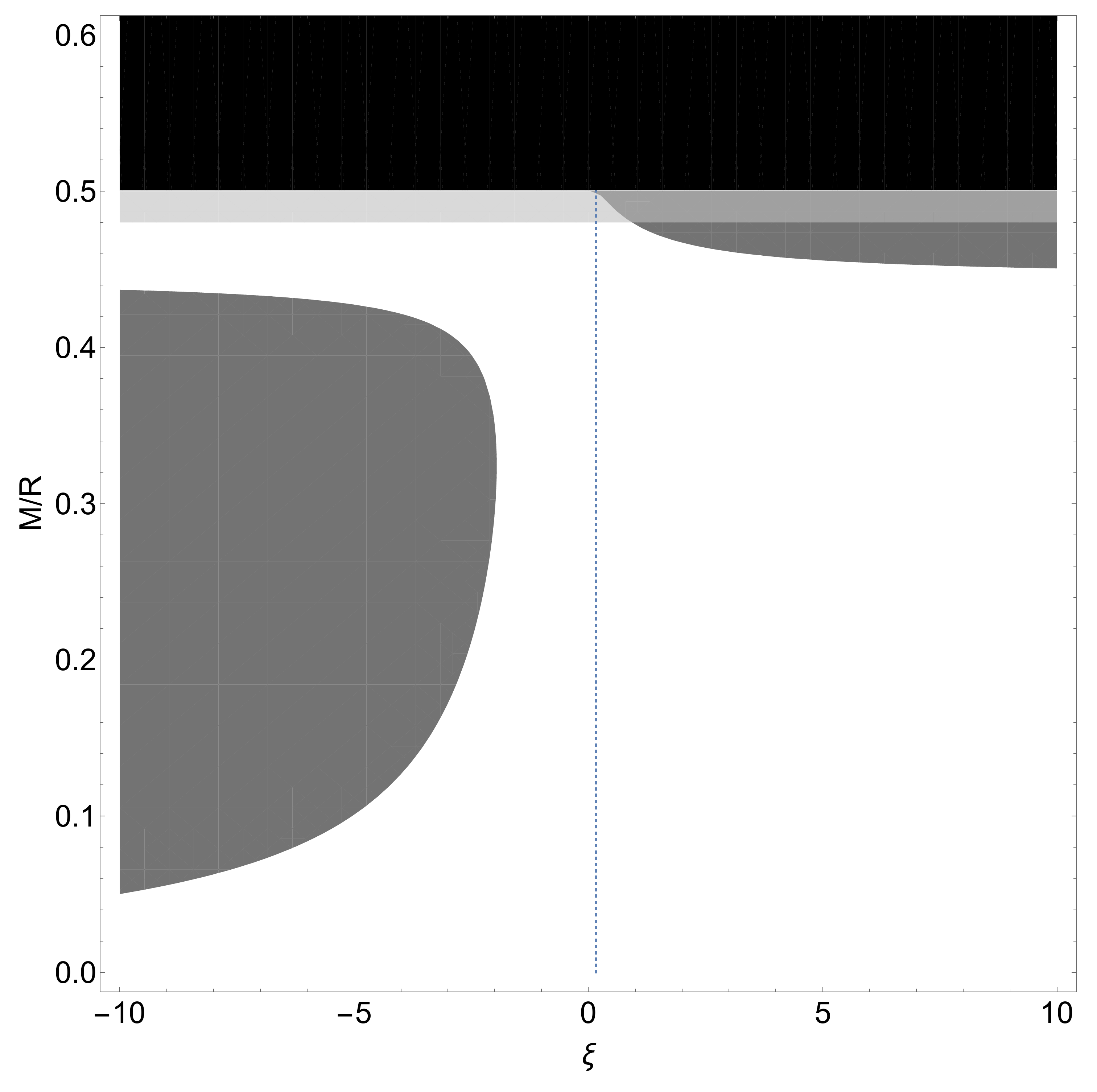}
\caption{ The black region is excluded because no static spherical 
shell can exist for $\mathsf{R}<\mathsf{R}_{2}
\equiv M+ \sqrt{M^2 +Q^2}$ when $|Q| \leq M$. 
The translucent gray area contains static shell configurations 
with $Q/M=0$ which violate the dominant energy condition.
The dark gray areas depict the regions where the 
``vacuum awakening effect" is triggered.  The vertical dotted 
line corresponds to the conformal coupling, $\xi=1/6$. It intercepts
the instability island inside the region where the dominant energy condition
is violated. This graph 
should be used as a benchmark for the sake of comparison with 
Figs.~\ref{fig3:QlessM_0_5}, \ref{fig4:QeqM}, and~\ref{fig5:MlessQ_1_5} 
where $Q/M\neq0$. 
}
\label{fig2:QlessM_0}
\end{figure}

\begin{figure}[htb]
\includegraphics[scale=0.25]{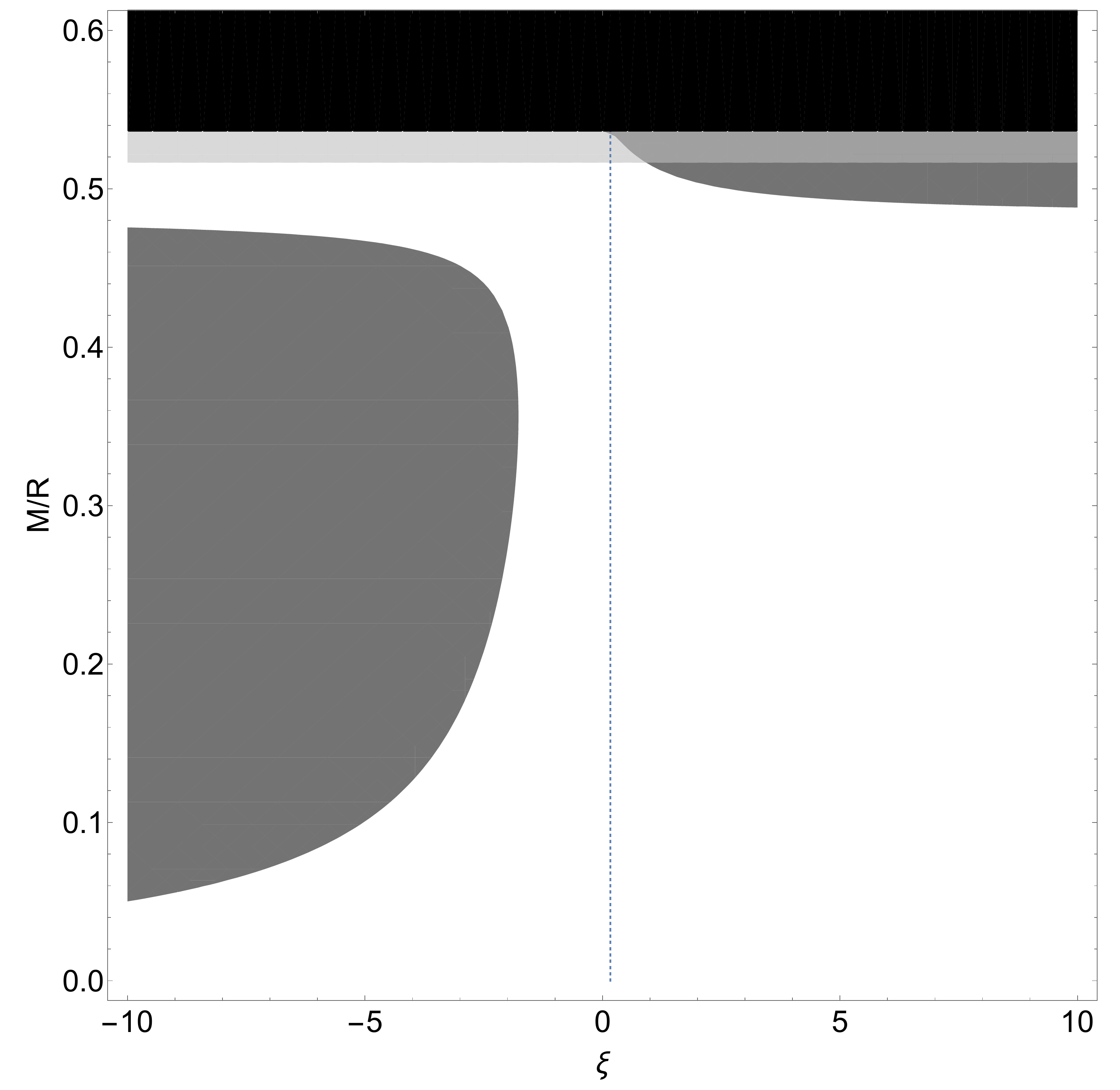}
\caption{This figure considers a shell with
$|Q|/M=0.5$. The same color convention  as in 
Fig.~\ref{fig2:QlessM_0} is assumed here.
}
\label{fig3:QlessM_0_5}
\end{figure}

\noindent {\bf Case $|Q| < M$:}
The general marginal solution  external to the shell for arbitrary $l$
can be written as~\cite{cm}
\begin{equation} \label{solution_outQ<M}
	\psi^+_{0 l} (r) = B^{(1)}_l r P_l \left( \frac{r-M}{\sqrt{M^2-Q^2}} \right)  
	                 + C^{(1)}_l r Q_l \left( \frac{r-M}{\sqrt{M^2-Q^2}} \right),
\end{equation}
where $B^{(1)}_l,C^{(1)}_l = {\rm const} \in \mathbb{R}$ guarantees 
that $\psi^+_{0 l}$ is a real function, $P_l(z)$ and $Q_l(z)$ are Legendre
functions of the first and second kinds~\cite{gr}, respectively, and we note that 
$Q_l[(r-M)/\sqrt{M^2-Q^2}]>0$ everywhere outside the shell.
Then, for $l=0$   
\begin{eqnarray} \label{solution_outQ<M,l=0}
	\psi^+_{0 0} (r) &=& B^{(1)}_0 r   
	                 + C^{(1)}_0 r \ln 
	                 \left( 
	                 \frac{r-M + \sqrt{M^2-Q^2}}{r-M - \sqrt{M^2-Q^2}} 
	                 \right)^{1/2}
\\
&\stackrel{r \gg M}{\sim}&
                    B^{(1)}_0 r   
	                 + C^{(1)}_0 \sqrt{M^2-Q^2}. 
\nonumber
\end{eqnarray}
Now, we impose the constraints~(\ref{continuity}) and~(\ref{discontinuity}) 
on the solutions~(\ref{solution_in,l=0}) and~(\ref{solution_outQ<M,l=0}) 
obtaining the following relationships among the integration constants:
\begin{eqnarray}\label{B10/A10}
	\frac{B^{(1)}_0}{A^{(1)}_0} &=& 1+ \frac{\xi\mathsf{R}/M}{\sqrt{1-Q^2/M^2}} 
	\left(
	-2+ \frac{3M}{ \mathsf{R}} -\frac{Q^2}{\mathsf{R}^2}
	+2 f(\mathsf{R})^{1/2} 
	\right)
	\nonumber \\
	&\times& \ln \left(\frac{-1+\mathsf{R}/M+\sqrt{1-Q^2/M^2}}{-1+\mathsf{R}/M-\sqrt{1-Q^2/M^2}   }\right)
\end{eqnarray}
and
\begin{equation}
\frac{C^{(1)}_0}{A^{(1)}_0} = 
\frac{2(1- B^{(1)}_0/A^{(1)}_0) }{\ln 
	                 \left[ 
	                 (\mathsf{R}-M + \sqrt{M^2-Q^2})/(\mathsf{R}-M - \sqrt{M^2-Q^2}) 
	                 \right]}.
\label{C10/A10}
\end{equation}
Next, we recall the discussion below Eq.~(\ref{aux4}) and impose 
$B^{(1)}_0/A^{(1)}_0=0$ to render the asymptotic form of the marginal 
solution as desired. This establishes a relationship between 
the coupling constant $\xi$ and the shell parameters $M/\mathsf{R}$, 
$Q/M$, which allows us to draw the boundary of the unstable regions in 
Figs.~\ref{fig2:QlessM_0} and~\ref{fig3:QlessM_0_5} for  $Q=0$ and 
$Q/M =0.5 $, respectively.

\begin{figure}[htb]
\includegraphics[scale=0.27]{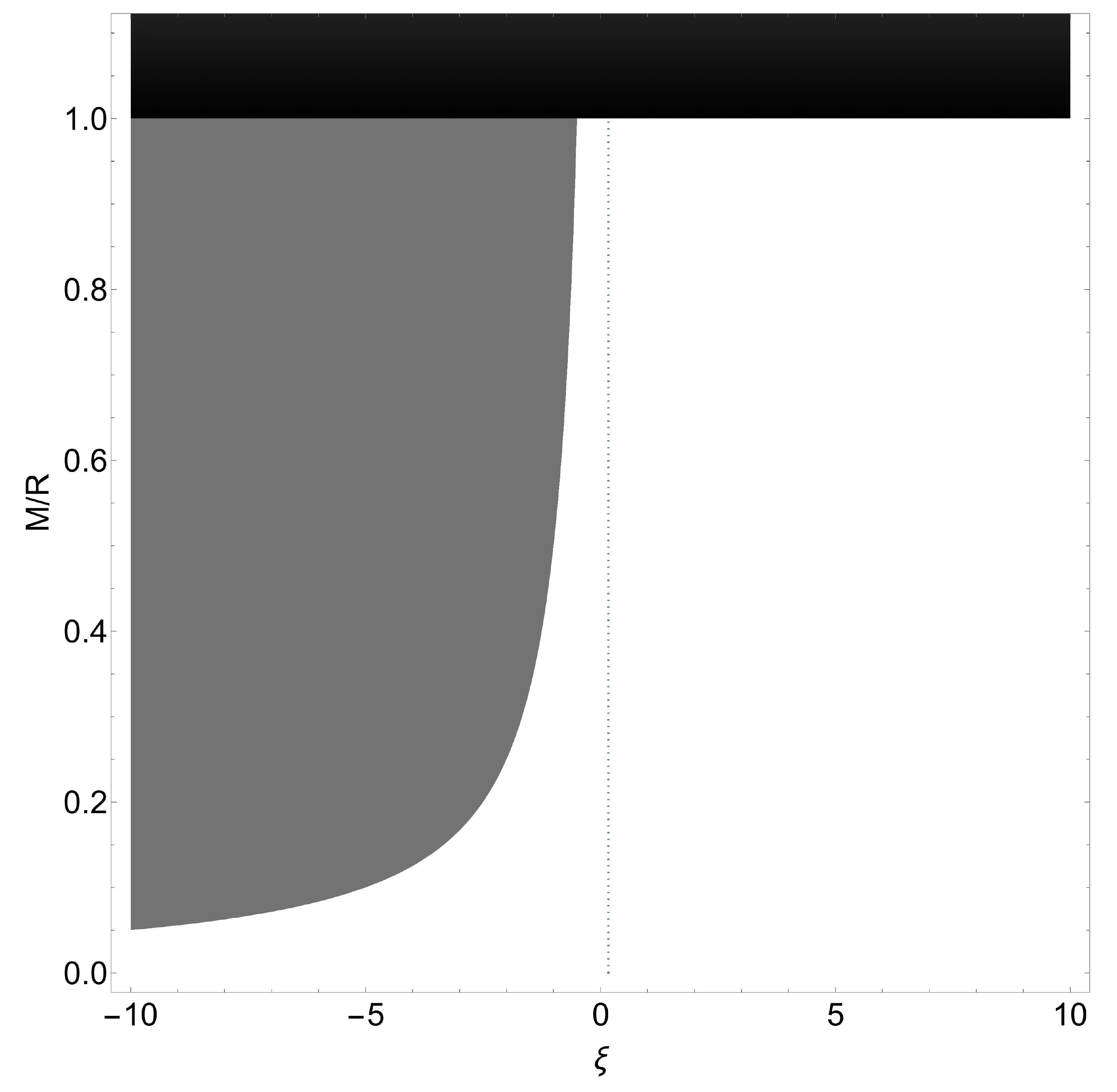}
\caption{ This figure considers a shell with
$|Q|/M=1$. The same color convention  as in 
Fig.~\ref{fig2:QlessM_0} is assumed here. 
We see that the instability islands which appear 
in Figs.~\ref{fig2:QlessM_0} and~\ref{fig3:QlessM_0_5} 
for $\xi>0$ are not present in this case. 
}
\label{fig4:QeqM}
\end{figure}

\noindent {\bf Case $|Q| = M$:}
Let us write the general marginal solution external to the shell in this case
as
\begin{equation} \label{solution_outQ=M}
	\psi^+_{0 l} (r) = B^{(2)}_l r \frac{(r-M)^l}{(\mathsf{R}-M)^l}   
	                 + C^{(2)}_l r \frac{(\mathsf{R}-M)^{l+1}}{(r-M)^{l+1}}	             
\end{equation}
with $B^{(2)}_l,C^{(2)}_l \in \mathbb{R}$ being constants.   
In the  particular case where $l=0$, we obtain 
\begin{equation} \label{solution_outQ=M,l=0}
	\psi^+_{0 0} (r) = B^{(2)}_0 r    
	                 + C^{(2)}_0 (\mathsf{R}-M) {r}/{(r-M)}.
\end{equation}
Now, by imposing the constraints~(\ref{continuity}) and~(\ref{discontinuity}) 
on Eqs.~(\ref{solution_in,l=0}) and~(\ref{solution_outQ=M,l=0}), we obtain
\begin{equation}\label{B20/A20}
	\frac{B^{(2)}_0}{A^{(2)}_0} = 1 + \frac{2 \xi M}{\mathsf{R}} 
	\;\;\;{\rm and} \;\;\; \frac{C^{(2)}_0}{A^{(2)}_0} 
	 = 1- \frac{B^{(2)}_0}{A^{(2)}_0}.
\end{equation}
The boundary of the unstable region in Fig.~\ref{fig4:QeqM} is obtained by demanding 
$B^{(2)}_0/A^{(2)}_0=0$.

\begin{figure}[htb]
\includegraphics[scale=0.25]{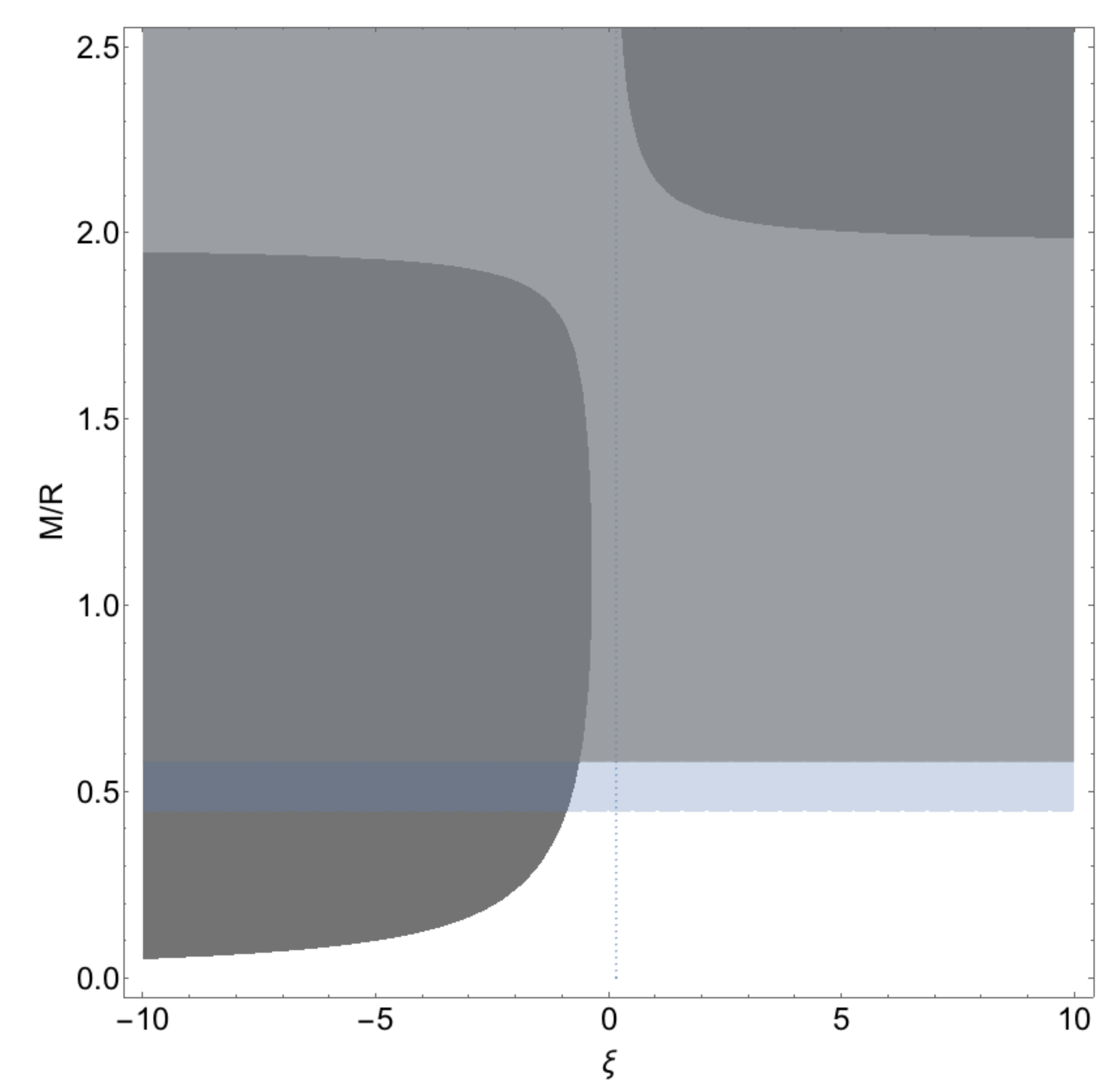}
\caption{This figure considers a shell with $|Q|/M=1.5$. 
The dark region is not present here in contrast to the 
previous three figures because for $|Q|>M$ there are 
static shell configurations for any $M/\mathsf{R}$. 
On the other hand, the energy conditions are 
typically more restrictive. The light 
translucent gray area corresponds to configurations where the 
strong energy condition is violated. The dark translucent 
gray region is excluded by the weak, strong, and dominant 
energy conditions.
}
\label{fig5:MlessQ_1_5}
\end{figure}

\noindent {\bf Case $|Q| > M$:} 
Finally, the general marginal solution external to the shell in this case
can be written as
\begin{eqnarray} \label{solution_outQ>M}
\psi^+_{0 l} (r) &=& B^{(3)}_l r (-i)^{l}P_l 
	                   \left( \frac{-i(r-M)}{\sqrt{Q^2-M^2}} \right)  
\nonumber \\
	               &+& C^{(3)}_l r (-i)^{l+1} Q_l 
	                   \left( \frac{-i(r-M)}{\sqrt{Q^2-M^2}} \right),
\end{eqnarray}
where $B^{(3)}_l,C^{(3)}_l \in \mathbb{R}$ are constants 
and  $\psi^+_{0 l}$ are
real functions. We then make use of Eq.~(8.834-2) of Ref.~\cite{gr}
to show that the Legendre functions of the second kind can be cast as
\begin{equation}
Q_l(-ix) =  P_l (-ix) Q_0(-ix)  - W_{l-1}(-ix), \;\;\;
x \in \mathbb{R},
\label{LegendreQ}
\end{equation}
where 
$$
W_{l-1} (-ix) = \sum_{k=1}^l \frac{P_{k-1} (-ix) P_{l-k} (-ix)}{k},\;\;\; 
W_{-1} (-ix) \equiv 0.
$$
Now, by choosing the branching cut for $\ln z$ along the line
$z= iy$, $y\geq 0$, we write
\begin{eqnarray}
Q_0(-ix) 
&=&
\frac{1}{2}\ln\left( \frac{-ix+1}{-ix-1}\right)
\nonumber \\
&=&i(\pi/2-\arctan x),
\end{eqnarray}
from which it can be seen that 
$$
(-i)^{l+1} Q_l  \left( \frac{-i(r-M)}{\sqrt{Q^2-M^2}} \right) >0.
$$
As before, we take $l=0$ in Eq.~(\ref{solution_outQ>M}), obtaining
\begin{eqnarray} \label{solution_outQ>M,l=0}
\psi^+_{0 0} (r)\!\!\!\!\!\!\!\!&=& \!\!\!\!\!\!\!\! B^{(3)}_0 r   
	                 +C^{(3)}_0  r \left[\frac{\pi}{2} -
	                  \arctan \left( \frac{r-M}{\sqrt{Q^2-M^2}}\right)\right]
	                \\
&\stackrel{r \gg Q}{\sim}&
                    B^{(3)}_0 r   
	                 + C^{(3)}_0 \sqrt{Q^2-M^2}.  
	                 \nonumber
\end{eqnarray}
Again, by imposing constraints~(\ref{continuity}) and~(\ref{discontinuity}) 
on Eqs.~(\ref{solution_in,l=0}) and~(\ref{solution_outQ>M,l=0}), we
obtain 
\begin{eqnarray}\label{B30/A30}
	\frac{B^{(3)}_0}{A^{(3)}_0} &=& 1 + \frac{2 \xi\mathsf{R}/M}{\sqrt{Q^2/M^2-1}} 
	\left(
	-2+ \frac{3M}{ \mathsf{R}} -\frac{Q^2}{\mathsf{R}^2}
	+2 f(\mathsf{R})^{1/2} 
	\right)
	\nonumber \\
	&\times& 
	\left[\frac{\pi}{2} -\arctan \left( 
	               \frac{-1+\mathsf{R}/M }{\sqrt{Q^2/M^2-1}} \right)\right]
\end{eqnarray}
and
\begin{equation}
\frac{C^{(3)}_0}{A^{(3)}_0} = 
\frac{1- B^{(3)}_0/A^{(3)}_0}{\pi/2-\arctan [(\mathsf{R}-M)/\sqrt{Q^2-M^2}]}.
\label{C30/A30}
\end{equation}
The boundary of the  unstable regions in Fig.~\ref{fig5:MlessQ_1_5} was drawn by
demanding  $B^{(3)}_0/A^{(3)}_0=0$.

In Figs.~\ref{fig2:QlessM_0}--\ref{fig5:MlessQ_1_5} we show the parameter-space 
region where tachyonic modes exist for $|Q|/M=0,\, 0.5,\, 1,\, 1.5$, rendering the 
scalar vacuum unstable. The black areas represent regions where there are 
no static shell configurations. For any fixed $|Q|/M$, there is a small enough 
$ (M/ \mathsf{R})_0$ such that for $ 0< M/ \mathsf{R}< (M/ \mathsf{R})_0$ the 
boundary of the unstable regions is determined by 
$\xi = -\mathsf{R}/(2 M) + {\cal{O}} (M/\mathsf{R})$
being independent of the value of $|Q|/M$. As $M/\mathsf{R}$ increases, the influence of
the charge on the instability regions becomes more noticeable. In particular, the
minimum value of $M/\mathsf{R}$ required for some scalar field with $\xi>0$ to become
unstable grows from $4/9$ when $Q=0$ to $1$ when $|Q|\to M_-$. Scalar fields with 
$\xi \geq 0$ are stable in the spacetime of shells with $|Q|=M$. We recall 
that for $|Q|>M$ there are static shell configurations for any $M/\mathsf{R}$ 
at the cost of having the weak, strong and dominant energy conditions being 
violated by shells with 
large enough $M/\mathsf{R}$. Finally, it is interesting to note that although 
there are shell configurations which allow for the existence of tachyonic modes 
for the conformal scalar field case, $\xi=1/6$, at least some energy condition 
will be violated by the corresponding shell.  
 
\section{Vacuum energy density }
\label{sec:vacuum}

Here, we analyze the exponential growth of the vacuum energy density
(as measured by static observers)
$\rho_V \equiv -\langle \hat{T}^0_{\;\; 0} 
          \rangle \equiv -\langle 0_{\rm in}| \hat{T}^0_{\;\; 0} 
|0_{\rm in} \rangle$ as a result of the instability induced by the 
appearance of tachyonic modes. We assume the vacuum $|0_{\rm in}\rangle$ 
to be the no-particle state as defined according to the oscillating modes
at the infinite past, where the initial shell parameters, $M_{\it in}$, 
$Q_{\it in}$, and $\mathsf{R}_{\it in}$, do not allow for 
the existence of tachyonic modes. Then, by assuming that the shell evolves to the 
static configuration, characterized by $M$, $Q$, and $\mathsf{R}$, 
which allows for the existence of tachyonic modes, we can use the 
general expression obtained in Ref.~\cite{lv} to 
write the leading contribution to the vacuum energy density
(see Ref.~\cite{llmv} for a more comprehensive discussion):
\begin{equation}\label{leading_term_erg_den0}
\rho_V 
=
-\langle \hat{T}^{0}_{\;\; 0}\rangle^- H(-\ell)
-
\langle \hat{T}^{0}_{\;\; 0}\rangle^+ H(\ell)
-
\langle \hat{T}^0_{\;\; 0}\rangle_{\cal S},
\end{equation}
where  $H(\ell)$ is the Heaviside step function (with $\ell$ being the proper 
distance of the geodesics which orthogonally intercept $\cal{S}$) and 
\begin{equation}\label{leading_term_erg_den-}
\langle \hat{T}^{0}_{\;\; 0}\rangle^-
\sim -\frac{\kappa}{8\pi}
\frac{e^{2\bar{\Omega}t} }{r^2} 
\frac{d}{dr} 
\left[ 
       \left( \frac{1-4\xi}{4\bar \Omega}  \right)  r^2
       \frac {d (\psi^-_{\bar\Omega 0 }(r)/r)^2}{dr}
       \right],      
\end{equation}
\begin{align}\label{leading_term_erg_den+}
\langle \hat{T}^{0}_{\;\; 0}\rangle^+
&
\sim -\frac{\kappa}{8\pi}   
\frac{e^{2\bar{\Omega}t}}{r^2} 
\frac{d}{dr} 
\left[ 
       \left( \frac{1-4\xi}{4\bar \Omega}  \right)r^2 f(r) 
       \frac {d (\psi^+_{\bar\Omega 0 }(r)/r)^2}{dr}
\right.       
\nonumber \\
&
       \left.         
       +\frac{\xi M}{{\bar{\Omega}}} \left( 1- \frac{Q^2}{M r} \right)
       \left( \frac{\psi^+_{\bar{\Omega} 0}(r)}{r} \right)^2 
       \right],
\end{align}
and
\begin{align}\label{leading_term_erg_denS}
\langle \hat{T}^{0}_{\;\; 0}\rangle_{\cal S}
&
\sim -\frac{\kappa}{8\pi} e^{2\bar{\Omega}t} 
\frac{f (\mathsf{R})^{-1/2}}{\mathsf{R}}
\frac{\xi}{\bar{\Omega}} \left[
              (1-4\xi) \left( 
                       -2+ \frac{3M}{\mathsf{R}}
                       \right.                       
\right.
\nonumber \\
&
\left.
\left.
       -\frac{Q^2}{\mathsf{R}^2} +2 f (\mathsf{R})^{1/2} 
       \right) 
              + \frac{M}{\mathsf{R}}-\frac{Q^2}{\mathsf{R}^2}
\right] 
\left(\frac{\psi^+_{\bar{\Omega} 0}(\mathsf{R}  )}{\mathsf{R}}\right)^2
\delta(\ell),
\end{align}
are the leading vacuum contributions to the energy density inside, outside, 
and on the shell, respectively. Here, $\kappa$ is a positive constant 
of order one (see Ref.~\cite{lv} for more details) and $\bar{\Omega}$ 
represents the largest $\Omega$ 
among all tachyonic solutions. We note that although the absolute value of 
the vacuum energy density increases exponentially in time, 
the growth is positive at some places and negative at some other ones
in such a way that the total vacuum energy is conserved~\cite{lv}.
Eventually, the spacetime must react to the vacuum energy growth
leading the whole system to evolve into some final stable configuration 
with no tachyonic modes. The analysis of the corresponding evolution in 
the context of semiclassical gravity is well known to be difficult. 
However, it was recently shown that the scalar field should
lose coherence fast enough to allow backreaction to be treated in the
much easier context of general relativity~\cite{llmv2}.

\section{Conclusions}
\label{sec:finalremarks}

We have analyzed the stability of a nonminimally coupled
free scalar field in the spacetime of charged spherical shells. 
The impact of the charge on the instability is enhanced
for more compact configurations. Also, the cases 
$|Q| \leq M$ and $|Q|>M$ differ  because in the latter case
there are static shell configurations for every radius $\mathsf{R}$. 
Notwithstanding, some of them may violate the weak, strong and dominant
energy conditions. The presence of charge does not alter the fact that 
spherically symmetric shells which are able to awake the vacuum for 
conformally coupled scalar fields, $\xi=1/6$, do not satisfy at least the 
dominant energy condition. Finally, we have calculated the 
expectation value of the vacuum energy density in order to make explicit
its exponential growth in time.

\acknowledgments

A.\ L., D.\ V., and J.\ S.\ were partially (A.\ L., D.\ V.) and 
fully (J.\ S.) supported by S\~ao Paulo Research Foundation 
(FAPESP) under Grants No.\ 2014/26307-8, No.\ 2013/12165-4, and 
No.\ 2013/07105-2 respectively, while W.\ L., R.\ M., and G.\ M.\
were fully (W.\ L., R.\ M.) and partially (G.\ M.) 
supported by Conselho Nacional de Desenvolvimento 
Cient\'\i fico e Tecnol\'ogico (CNPq).

\end{document}